\documentclass[11pt]{article}
\usepackage[a4paper,margin=25mm]{geometry}
\usepackage[T1]{fontenc}
\usepackage{lmodern}
\usepackage{amsmath,amssymb}
\usepackage{booktabs,array,tabularx}
\usepackage{graphicx}
\usepackage{microtype}
\usepackage[authoryear,round]{natbib}
\setcitestyle{aysep={,},citesep={;}}
\usepackage[hidelinks]{hyperref}
\newcommand{\model}{X-Embedding}

\newcommand{\norm}[1]{\left\lVert #1\right\rVert_2}
\title{\model: Two-Stage Training for Multilingual Patent Prior-Art Retrieval}
\author{\textbf{Wentao Wu, Xu Wang, Linqing Chen\textsuperscript{*},}\\[3pt]
\textbf{Fenhong Zhu, Hanmeng Zhong, Yuanxiang Huangfu}\\[4pt]
Patsnap\\[3pt]
\small\texttt{\{wuwentao3, wangxu2, chenlinqing,}\\[1pt]
\small\texttt{zhufenhong, zhonghanmeng, huangfuyuanxiang\}@patsnap.com}\\[3pt]
\small\textsuperscript{*}Corresponding author: Linqing Chen}
\date{}

\usepackage[table]{xcolor}
\usepackage{tikz}
\usetikzlibrary{arrows.meta,positioning,calc}
\usepackage{caption}
\usepackage{titlesec}
\usepackage{titling}
\usepackage{placeins}
\definecolor{StageBlue}{HTML}{304F70}
\definecolor{StageTeal}{HTML}{286765}
\definecolor{RuleGray}{HTML}{65717D}
\definecolor{TableTint}{gray}{0.95}
\titleformat{\section}{\large\bfseries}{\thesection}{0.65em}{}
\titleformat{\subsection}{\normalsize\bfseries}{\thesubsection}{0.65em}{}
\titlespacing*{\section}{0pt}{2.2ex plus .4ex minus .2ex}{1.1ex}
\titlespacing*{\subsection}{0pt}{1.7ex plus .3ex minus .2ex}{0.7ex}
\pretitle{\begin{center}\LARGE\bfseries}
\posttitle{\par\end{center}\vspace{2mm}}
\preauthor{\begin{center}\vspace{-6mm}\normalsize}
\postauthor{\par\end{center}}
\predate{\begin{center}\small}
\postdate{\par\end{center}\vspace{-16mm}}
\begin{document}
\maketitle
\begin{abstract}
Multilingual patent prior-art retrieval requires identifying X-category references that are particularly relevant to the novelty or inventive step of a claimed invention, beyond topical similarity across languages and document views. We present \model, a two-stage framework that first learns patent-domain representations and then optimizes X-category retrieval relevance. The first stage aligns complementary sections within patents and corresponding sections across different-language patent-family members, learning from both patent structure and multilingual expressions. The second stage learns X-category relevance using examination-report X citations as positives and negatives drawn from retrieval candidates and A-category citations. Matryoshka representation learning (MRL) enables 64- and 1024-dimensional embeddings within one model. Evaluation on a common Chinese, English, and Japanese collection with four query views yields 89.09\% overall macro Hit@100 at 64 dimensions, exceeding Qwen3-Embedding-0.6B and RoBERTa by 21.25 and 3.68 percentage points, respectively. Under the same second-stage training and evaluation conditions, two-stage training outperforms direct supervised fine-tuning by 3.19 points, with gains of 5.34 and 4.72 points for English and Japanese and higher scores in nine of twelve language--view combinations. Overall macro Hit@10 and Hit@20 also improve by 3.71 and 3.79 points, respectively. These results support sequential patent-domain and task adaptation for improving aggregate X-category retrieval on this multilingual collection.
\end{abstract}

\section{Introduction}
\label{sec:introduction}
Patent prior-art retrieval seeks earlier disclosures relevant to assessing the novelty or inventive step of a claimed invention. Its relevance criterion extends beyond shared terminology or membership in the same technological field: two patents may describe similar components yet differ in the connections, operating conditions, or sequence of steps that define the invention. We focus on X-category references, which are particularly relevant when considered individually under the search-report convention \citep{uspto_xcategory}. Retrieving these references requires a representation that connects the technical content of a query patent to examination-relevant disclosures, even when their wording and presentation differ.

Patent retrieval combines lexical and semantic evidence in several ways. BM25 provides exact-term matching \citep{bm25}, dense retrievers learn correspondences beyond shared vocabulary, and hybrid retrieval and reranking combine or refine these signals \citep{yousefiramandi2026benchmarking}. Exact-term matching is valuable for specialized terminology but cannot by itself resolve synonymous or different-language expressions. Learned semantic similarity addresses a broader range of expressions, yet proximity in an embedding space must still reflect the intended relevance criterion. For X-category retrieval, the central representation-learning problem is to distinguish examination-relevant disclosures from patents that are merely close in topic.

Domain-specific training provides a means of connecting representations to patent evidence. PaECTER adapts an encoder using examiner-added citations \citep{ghosh2024paecter}, while patent-section self-supervision learns from related views within a document \citep{zuo2025selfsupervision}. QaECTER jointly learns citation retrieval and multi-view self-alignment \citep{djemmal2026qaecter}. \citet{yousefiramandi2026benchmarking} further study cross-section training and mixtures of taxonomy, citation, and cross-section pairs, including adaptation of Qwen3-Embedding-0.6B. Building on these studies, we investigate within-patent cross-section and within-family cross-language correspondence as a preliminary adaptation stage followed by X-citation relevance learning, assessing its benefit relative to direct supervised fine-tuning.

The multilingual setting introduces variation in both document view and language. An abstract summarizes an invention, claims specify the subject matter for which protection is sought, and a description develops the technical disclosure in greater detail. The TAC view concatenates the title, abstract, and claim fields. Such differences in length, emphasis, and specificity are reflected in patent benchmarks that evaluate multiple query types and asymmetric text representations \citep{djemmal2026qaecter,ayaou2025patenteb}. Chinese, English, and Japanese expressions introduce further variation. Within-patent sections and different-language patent-family members therefore provide complementary opportunities to learn correspondence despite changes in how technical content is presented.

The distinction between these correspondences and X-category relevance motivates a sequential approach. Sections within a patent and corresponding sections across its family members provide weak supervision for representing related technical content. Examination-report X citations provide task supervision for ranking relevant prior-art references above competing candidates. Our hypothesis is that the former can provide a useful initialization for learning the latter. Specifically, we ask whether cross-section and multilingual family alignment before X-category supervision improves retrieval over direct supervised fine-tuning under the same subsequent training conditions.

We develop \model\ on Qwen3-Embedding-0.6B \citep{qwen3embedding} to examine this hypothesis. Stage I uses abstract--claim and claim--description pairs within patents, together with abstract--abstract and claim--claim pairs across different-language members of the same family. Stage II adapts the resulting encoder to X-category retrieval using examination-report X citations and selected negative candidates. Domain alignment thus precedes citation-supervised relevance learning. We assess this training sequence through comparisons with the base encoder, the intermediate Stage I model, and a direct supervised fine-tuning control with matched second-stage settings. Evaluation across query languages and views characterizes where the additional domain-adaptation stage improves retrieval.

Our contributions are twofold:
\begin{enumerate}
\item \textbf{Sequential domain adaptation for multilingual X-category retrieval.} Stage I jointly learns cross-section correspondence within patents and same-section correspondence across different-language family members; Stage II learns retrieval relevance from examination-report X citations. The framework uses these two forms of patent-content correspondence to initialize task-supervised training, connecting multilingual patent representation learning to the X-category retrieval objective.
\item \textbf{An empirical analysis of the aggregate benefit of preliminary adaptation and its language--view variation.} Under the same second-stage training and evaluation conditions, the complete model improves overall macro Hit@100 by 3.19 percentage points over direct supervised fine-tuning at 64 dimensions, with gains in nine of twelve language--view combinations. Hit@100 gains are concentrated in English and Japanese, while Hit@10 and Hit@20 improve in all three query-language groups, characterizing the benefit across languages, views, and retrieval cutoffs.
\end{enumerate}

\section{Related Work}
\label{sec:related}
\subsection{Patent representations and retrieval relevance}
Patent representation learning addresses several objectives, including technological similarity, classification, and prior-art relevance. PatentSBERTa studies technological distance and classification \citep{patentsberta}, while PatentMatch constructs claim--passage matching examples from examiner citations \citep{patentmatch}. PaECTER adapts BERT for Patents using examiner-added citation supervision \citep{ghosh2024paecter}. Graph-based patent retrieval further represents invention features and their relationships explicitly and learns from examiner citations \citep{graphpatent}. These studies learn patent representations from technological similarity, localized text matching, and citation relationships.

These tasks differ in both the unit of supervision and the meaning of relevance. Technological similarity concerns related inventions, claim--passage matching concerns localized textual evidence, and document retrieval ranks complete candidate references. Consequently, performance on one task does not directly characterize performance on another. We construct retrieval relevance labels from X-category citations recorded in examination reports.

\subsection{Citation supervision and multi-view alignment}
\citet{zuo2025selfsupervision} learn patent representations through self-supervision using document sections as related views. QaECTER jointly trains citation retrieval and multi-view self-alignment using X-, Y-, and A-category citation pairs and views of the same patent \citep{djemmal2026qaecter}. Sophia-bench evaluates twelve query representations against title--abstract--claims--description (TACD) candidates. Its relevance pool includes cited and citing documents in the XY and A categories; evaluation uses English text, including translated patent-family representatives.

\subsection{Patent embedding evaluation and multi-stage training}
PatenTEB broadens patent embedding evaluation across task types and asymmetric query--document representations \citep{ayaou2025patenteb}. DAPFAM studies family-level retrieval and cross-domain transfer \citep{ayaou2025dapfam}. \citet{yousefiramandi2026benchmarking} adapt encoders including Qwen3-Embedding-0.6B: R3 pairs each patent abstract with its first independent claim, whereas R4 pools taxonomy, citation, and cross-section pairs. Their study includes a data-volume-matched control and examines supervision across retrieval, classification, and clustering. These studies highlight the importance of specifying the relevance unit, query view, domain, and representation size.

Multi-stage embedding training has also been explored in general-purpose retrieval. GTE combines unsupervised pretraining with supervised adaptation \citep{li2023gte}.

Our approach assigns patent-content correspondence and X-category relevance to two successive training stages. Whereas QaECTER jointly trains citation retrieval and multi-view alignment \citep{djemmal2026qaecter} and R4 mixes pairs from different sources \citep{yousefiramandi2026benchmarking}, we first combine cross-section pairs with same-section pairs from different-language family members, then train for retrieval using X-category citations. The distinction concerns the organization of supervision and the retrieval target. A direct-supervision control with matched second-stage settings evaluates the aggregate benefit of adding patent-domain adaptation before X-category supervision.

Our evaluation uses X-category relevance labels, Chinese, English, and Japanese queries, and TAC candidates, whereas Sophia-bench uses broader citation-based relevance labels, English queries, and TACD candidates \citep{djemmal2026qaecter}. Our empirical conclusions follow from model comparisons on a common evaluation collection, rather than comparisons with scores reported by these studies under different collections and relevance definitions.

\section{Background}
\label{sec:background}
\subsection{X-category patent retrieval}
\label{sec:feature-coverage}
This study defines patent prior-art retrieval as follows: given a query patent, retrieve X-category patents from a candidate corpus. We use the X category of PCT international search reports, as described in the USPTO examination manual: an X reference is particularly relevant when taken alone and may prejudice novelty or inventive step, with common general knowledge considered where applicable \citep{uspto_xcategory}. Training positives are X-category citations recorded in patent examination search reports.

For novelty, an X-category patent must disclose the necessary claim elements together in the arrangement required by the claim \citep{uspto_anticipation}. The relevant technical disclosure includes the technical objects and components; their connections, relative positions, and functional interactions; numerical and geometric restrictions; ordered method steps; and the distinguishing feature combination of the claimed invention.

\subsection{Multilingual, multi-view retrieval}
\label{sec:task}
Let $p$ be a source patent and let $\mathcal{V}=\{\mathrm{ABS},\mathrm{CLAIM},\mathrm{DESC},\mathrm{TAC}\}$ denote the query views. A query is $q_v=\phi_v(p)$ for $v\in\mathcal{V}$, while a candidate patent $d$ is represented by $t(d)=\phi_{\mathrm{TAC}}(d)$. Each view constitutes a separate query. The target set $\mathcal{X}(p)$ comprises the recorded X-category references for $p$ and is shared across its query views.

Given a candidate corpus $\mathcal{D}$, a dense retriever assigns a score $s_\theta(q_v,t(d))$ and returns
\begin{equation}
\mathcal{R}_K(q_v)=\operatorname{TopK}_{d\in\mathcal{D}}\,s_\theta(q_v,t(d)).
\label{eq:retrieval}
\end{equation}
Training seeks to assign higher scores to members of $\mathcal{X}(p)$. The encoder supports Chinese, English, and Japanese text, and evaluation is stratified by query language and view.

The query view determines which part of the source patent is available to the encoder, while the relevance target remains the recorded X-category reference set. Thus, an abstract query tests retrieval from a concise account of the invention, whereas a claim or description query supplies a different scope of technical disclosure. Evaluating these views separately characterizes retrieval from the information actually present in each input, rather than assuming that all sections convey equivalent evidence.

\section{Method}
\label{sec:methods}
\subsection{Framework overview}
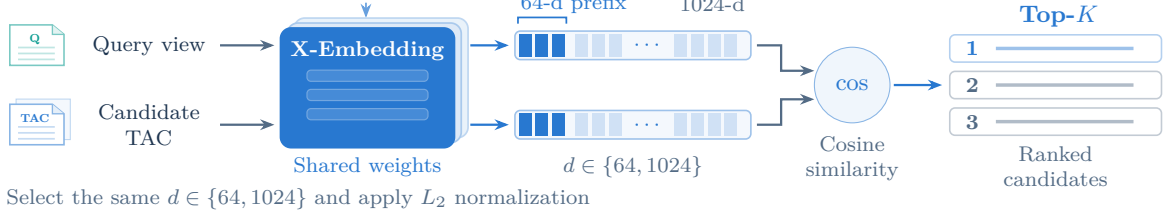
\begin{figure}[!htbp]
\centering
\definecolor{FigInk}{HTML}{1E3550}
\definecolor{FigWeak}{HTML}{0D9D92}
\definecolor{FigTask}{HTML}{2878D0}
\definecolor{FigAmber}{HTML}{D87B1D}
\definecolor{FigMuted}{HTML}{536B85}
\definecolor{FigLine}{HTML}{CCDDF0}
\newcommand{\patentglyph}[4]{%
  \begin{scope}[shift={(#1,#2)}]
    \path[fill=white,draw=#4!55,line width=.6pt]
      (0,0)--(.88,0)--(1.12,-.24)--(1.12,-1.02)--(0,-1.02)--cycle;
    \path[fill=#4!12,draw=#4!55,line width=.5pt]
      (.88,0)--(.88,-.24)--(1.12,-.24);
    \node[font=\scriptsize\bfseries,text=#4,scale=.85,transform shape] at (.52,-.35) {#3};
    \foreach \y in {-.57,-.72,-.87}
      \draw[#4!26,line width=.65pt] (.17,\y)--(.93,\y);
  \end{scope}}
\begin{tikzpicture}[x=1cm,y=.82cm,
  every node/.style={font=\footnotesize,align=center,inner sep=2pt,text=FigInk},
  flow/.style={-{Stealth[length=1.9mm]},line width=.85pt,draw=FigMuted},
  minor/.style={draw=FigLine,line width=.55pt},
  card/.style={rounded corners=4pt,line width=.65pt}]
% The figure follows a numbered learning path above a separate retrieval lane.
\node[anchor=west,font=\small\bfseries] at (0,.25) {A  Two-stage training};
\node[anchor=east,font=\scriptsize,text=FigMuted] at (15.4,.25) {Base encoder: Qwen3-Embedding-0.6B};
\path[card,fill=FigWeak!3,draw=FigWeak!25] (0,-.15) rectangle (7.05,-5.22);
\path[card,fill=FigTask!3,draw=FigTask!25] (8.35,-.15) rectangle (15.4,-5.22);
\draw[FigWeak,line width=2.0pt] (.22,-.15)--(6.83,-.15);
\draw[FigTask,line width=2.0pt] (8.57,-.15)--(15.18,-.15);
% Number badges and left-aligned titles supply a clear typographic hierarchy.
\node[fill=FigWeak,text=white,rounded corners=2pt,
  minimum size=.52cm,font=\footnotesize\bfseries] at (.49,-.63) {01};
\node[anchor=west,font=\small\bfseries,text=FigWeak] at (.94,-.63) {Stage I: Weak supervision};
\node[fill=FigTask,text=white,rounded corners=2pt,
  minimum size=.52cm,font=\footnotesize\bfseries] at (8.84,-.63) {02};
\node[anchor=west,font=\small\bfseries,text=FigTask] at (9.29,-.63) {Stage II: X-category supervision};
\draw[minor] (.3,-1.04)--(6.75,-1.04);
\draw[minor] (8.65,-1.04)--(15.10,-1.04);
% Section correspondence is rendered as three patent pages.
\node[anchor=west,text=FigMuted] at (.32,-1.38) {Same patent \textperiodcentered{} Different sections};
\patentglyph{.67}{-1.76}{ABS}{FigWeak}
\patentglyph{2.96}{-1.76}{CLAIM}{FigWeak}
\patentglyph{5.25}{-1.76}{DESC}{FigWeak}
\draw[flow,draw=FigWeak!70] (1.86,-2.27)--(2.86,-2.27);
\draw[flow,draw=FigWeak!70] (4.15,-2.27)--(5.15,-2.27);
% Family pairs retain section identity while the language badges differ.
\node[anchor=west,text=FigMuted] at (.32,-3.12) {Same family \textperiodcentered{} Different languages};
\path[card,fill=white,draw=FigLine] (.32,-3.47) rectangle (3.36,-4.55);
\path[card,fill=white,draw=FigLine] (3.69,-3.47) rectangle (6.73,-4.55);
\foreach \x/\field in {.53/ABS,3.90/CLAIM}{
  \begin{scope}[shift={(\x,-3.62)},scale=.67]
    \patentglyph{0}{0}{}{FigWeak}
    \patentglyph{2.74}{0}{}{FigTask}
  \end{scope}
  \node[fill=white,text=FigWeak,font=\scriptsize\bfseries,inner xsep=5pt]
    at (\x+1.28,-3.47) {\field};
  \draw[FigMuted!65,line width=.7pt] (\x+.83,-4.02)--(\x+1.74,-4.02);
  \fill[FigWeak!60] (\x+.83,-4.02) circle (1pt);
  \fill[FigTask!60] (\x+1.74,-4.02) circle (1pt);
  \node[fill=FigWeak!12,text=FigWeak,rounded corners=1pt,font=\scriptsize,
    inner xsep=3pt,inner ysep=1pt] at (\x+.37,-4.41) {$L_1$};
  \node[fill=FigTask!12,text=FigTask,rounded corners=1pt,font=\scriptsize,
    inner xsep=3pt,inner ysep=1pt] at (\x+2.21,-4.41) {$L_2$};
}
\node[fill=FigWeak!11,text=FigWeak,rounded corners=3pt,
  minimum width=6.38cm,minimum height=.40cm,font=\footnotesize\bfseries]
  at (3.525,-4.88) {InfoNCE: patent-view alignment};
% Only learned parameters, rather than examples, move between the stages.
\draw[flow,draw=FigWeak,line width=1.1pt] (7.05,-2.66)--(8.35,-2.66);
\node[text=FigWeak,font=\scriptsize,fill=white,inner sep=1pt] at (7.70,-2.34) {Transfer};
% A query is paired with an examiner-labelled positive and eight negatives.
\patentglyph{8.87}{-2.04}{Q}{FigWeak}
\node[font=\footnotesize\bfseries,text=FigWeak] at (9.43,-3.38) {Query view};
\node[font=\scriptsize,text=FigMuted,text width=2.0cm] at (9.43,-4.14)
  {ABS / CLAIM\\DESC / TAC};
\patentglyph{12.00}{-1.45}{X}{FigTask}
\node[text=FigTask,font=\footnotesize\bfseries,text width=2.0cm]
  at (14.10,-1.95) {X-citation\\positive};
\begin{scope}[shift={(12.0,-3.25)}]
  \path[fill=FigAmber!6,draw=FigAmber!35] (.18,.18) rectangle (1.30,-.84);
  \path[fill=FigAmber!7,draw=FigAmber!40] (.09,.09) rectangle (1.21,-.93);
  \patentglyph{0}{0}{TAC}{FigAmber}
\end{scope}
\node[text=FigAmber,font=\footnotesize\bfseries,text width=2.0cm]
  at (14.10,-3.76) {8 selected\\negatives};
\draw[flow,draw=FigTask,line width=1pt] (10.08,-2.47)--(11.82,-1.98);
\draw[flow,draw=FigAmber,densely dashed,line width=1pt] (10.08,-2.74)--(11.82,-3.71);
\node[circle,fill=white,draw=FigTask!60,text=FigTask,inner sep=1.2pt,
  font=\scriptsize] at (10.97,-2.22) {$+$};
\node[circle,fill=white,draw=FigAmber!60,text=FigAmber,inner sep=1.2pt,
  font=\scriptsize] at (10.97,-3.23) {$-$};
\node[fill=FigTask!11,text=FigTask,rounded corners=3pt,
  minimum width=6.38cm,minimum height=.40cm,font=\footnotesize\bfseries]
  at (11.875,-4.88) {InfoNCE + MRL (64 / 1024)};
% The retrieval lane uses the trained shared encoder, with no model duplication.
\node[anchor=west,font=\small\bfseries] at (0,-5.90) {B  Shared encoding and retrieval};
\draw[flow,draw=FigTask!70,densely dashed,line width=.65pt]
  (11.875,-5.22)--(11.875,-5.56)--(6.40,-5.56)--(6.40,-6.17)--(4.83,-6.17)--(4.83,-6.52);
\node[font=\scriptsize,text=FigTask,fill=white,inner sep=2pt]
  at (9.40,-5.56) {Trained weights};
% Small documents replace generic input boxes.
\begin{scope}[shift={(.12,-6.63)},scale=.64]
  \patentglyph{0}{0}{Q}{FigWeak}
\end{scope}
\node[anchor=west,font=\scriptsize,text width=1.85cm] at (.96,-6.96) {Query view};
\begin{scope}[shift={(.12,-7.91)},scale=.64]
  \path[fill=FigTask!5,draw=FigTask!30] (.13,.13) rectangle (1.25,-.89);
  \patentglyph{0}{0}{TAC}{FigTask}
\end{scope}
\node[anchor=west,font=\scriptsize,text width=1.85cm] at (.96,-8.24) {Candidate TAC};
\draw[flow] (2.90,-6.96)--(3.62,-6.96);
\draw[flow] (2.90,-8.24)--(3.62,-8.24);
% Layered slabs symbolize the encoder, not a count of architecture layers.
\path[card,fill=FigTask!10,draw=FigTask!30] (3.88,-6.52) rectangle (6.18,-8.45);
\path[card,fill=FigTask!16,draw=FigTask!40] (3.79,-6.60) rectangle (6.09,-8.53);
\path[card,fill=FigTask,draw=FigTask] (3.70,-6.68) rectangle (6.00,-8.61);
\node[text=white,font=\footnotesize\bfseries,scale=.9,transform shape] at (4.85,-7.04) {X-Embedding};
\foreach \y in {-7.45,-7.76,-8.07}{
  \path[rounded corners=1.5pt,fill=white!14!FigTask,draw=white!35!FigTask,
    line width=.45pt] (4.05,\y+.09) rectangle (5.65,\y-.09);
}
\node[text=FigTask,font=\scriptsize] at (4.85,-8.91) {Shared weights};
\draw[flow,draw=FigTask] (6.21,-6.96)--(6.72,-6.96);
\draw[flow,draw=FigTask] (6.21,-8.24)--(6.72,-8.24);
% Matryoshka prefixes: full vector outlines enclose highlighted leading entries.
\foreach \y in {-6.96,-8.24}{
  \path[rounded corners=2pt,fill=FigTask!5,draw=FigTask!45,line width=.6pt]
    (6.80,\y-.23) rectangle (9.92,\y+.23);
  \foreach \x in {6.85,7.07,7.29}{
    \fill[FigTask] (\x,\y-.18) rectangle (\x+.17,\y+.18);
  }
  \foreach \x in {7.60,7.82,8.04,8.91,9.13,9.35,9.57}{
    \fill[FigTask!22] (\x,\y-.18) rectangle (\x+.17,\y+.18);
  }
  \node[text=FigTask,font=\scriptsize] at (8.57,\y) {$\cdots$};
}
\draw[FigTask,line width=.65pt] (6.84,-6.61)--(6.84,-6.51)--(7.48,-6.51)--(7.48,-6.61);
\node[font=\scriptsize,text=FigTask,anchor=west] at (6.80,-6.29) {64-d prefix};
\node[font=\scriptsize,text=FigMuted,anchor=east] at (9.92,-6.29) {1024-d};
\node[text=FigMuted,font=\scriptsize] at (8.36,-8.91) {$d\in\{64,1024\}$};
% Similarity is shown as an operator; output is a ranked list, not a text box.
\node[circle,draw=FigTask!55,fill=FigTask!7,minimum size=1.0cm,
  font=\footnotesize\bfseries,text=FigTask] (score) at (11.26,-7.60) {$\cos$};
\draw[flow] (10.00,-6.96)--(10.32,-6.96)--(10.32,-7.37)--(10.76,-7.37);
\draw[flow] (10.00,-8.24)--(10.32,-8.24)--(10.32,-7.83)--(10.76,-7.83);
\node[text=FigMuted,font=\scriptsize] at (11.26,-8.78) {Cosine\\similarity};
\draw[flow,draw=FigTask] (11.81,-7.60)--(12.45,-7.60);
\node[text=FigTask,font=\footnotesize\bfseries] at (13.99,-6.49) {Top-$K$};
\foreach \rank/\y/\col in {1/-7.01/FigTask,2/-7.60/FigMuted,3/-8.19/FigMuted}{
  \path[rounded corners=2pt,fill=white,draw=\col!35,line width=.6pt]
    (12.55,\y-.22) rectangle (15.35,\y+.22);
  \node[text=\col,font=\scriptsize\bfseries] at (12.85,\y) {\rank};
  \draw[\col!45,line width=1.3pt] (13.16,\y)--(14.98,\y);
}
\node[text=FigMuted,font=\scriptsize] at (13.95,-8.91) {Ranked\\candidates};
\node[text=FigMuted,font=\scriptsize,anchor=west] at (0,-9.44) {Select the same $d\in\{64,1024\}$ and apply $L_2$ normalization};
\end{tikzpicture}
\caption{Two-stage training and retrieval in X-Embedding. (A) Stage I aligns sections within a patent and languages within a family ($L_1\ne L_2$); Stage II uses X-category supervision and MRL. (B) Queries and candidates are encoded independently with shared weights. Retrieval uses normalized representations of the same dimensionality: 64-dimensional prefixes (dark) or full 1024-dimensional embeddings. All experiments use 64 dimensions.}
\label{fig:pipeline}
\end{figure}

\model\ uses a shared Qwen3-Embedding-0.6B encoder \citep{qwen3embedding}. Queries and candidate texts are encoded independently into a common vector space. Training proceeds sequentially: Stage I learns from weakly supervised patent-view pairs, and Stage II adapts the encoder using task relevance labels. Matryoshka representation learning (MRL) \citep{mrl} trains nested representations for retrieval at 64 or 1024 dimensions. Figure~\ref{fig:pipeline} summarizes the training procedure and embedding construction.

\subsection{Stage I: weakly supervised patent-view alignment}
\label{sec:stage-one}
The first stage combines two sources of weak supervision. \emph{Within-patent cross-section alignment} uses ABS--CLAIM and CLAIM--DESC pairs drawn from the same patent. These pairs link a concise summary to the claimed subject matter and connect the claims to the detailed disclosure, aligning sections that differ in purpose and level of detail. \emph{Within-family cross-language alignment} uses ABS--ABS and CLAIM--CLAIM pairs drawn from different-language members of the same patent family. Pairing the same section type across languages is intended to teach the encoder how corresponding patent content is expressed in different languages. Family membership provides weak supervision for this correspondence. 

Stage I uses approximately 25.3 million weakly supervised patent-text pairs.

Let $\mathcal{W}$ denote the weak-pair dataset containing both within-patent cross-section pairs and within-family cross-language pairs. For a pair $(a_i,b_i)\in\mathcal{W}$ and a batch $\mathcal{B}$, an InfoNCE contrastive alignment loss \citep{infonce} takes the form
\begin{equation}
\ell_i^{\mathrm{view}}=-\log
\frac{\exp(s_\theta(a_i,b_i)/\tau)}
{\sum_{j\in\mathcal{B}}\exp(s_\theta(a_i,b_j)/\tau)},
\label{eq:weak}
\end{equation}
where $s_\theta$ is normalized embedding similarity and $\tau>0$ is a temperature. Other paired texts in the batch provide comparison candidates. Optimizing this objective draws paired texts closer in the representation space. The resulting parameters initialize the supervised stage.

\subsection{Stage II: supervised prior-art retrieval}
\label{sec:stage-two}
The second stage uses training examples $(q_i,d_i^+,\widetilde{\mathcal{N}}_i)$, where $q_i$ is a query view, $d_i^+$ is an X-category citation recorded in an examination search report, and $|\widetilde{\mathcal{N}}_i|=8$. Negatives are drawn from retrieval candidates and A-category citations through score-stratified sampling followed by language-prioritized selection. Candidate patents are represented by TAC.

Training queries must have at least one X-category positive and exactly 2,000 candidates before filtering. Retrieval uses ABS, CLAIM, DESC, or TAC queries, with the filing date as the upper bound on candidate publication dates. Candidates are deduplicated by patent identifier and simple family, excluding the query patent, X-category positives and their family members, and documents with invalid vector scores. A-category citations supply additional negatives: under the search-report convention, they describe the state of the art without prejudicing novelty or inventive step \citep{uspto_xcategory}.

Candidates are stratified jointly by vector score and BM25 rank, with high, medium, and low levels on each axis. Sampling up to four documents from each of the nine strata yields at most 36 retrieved negatives. The four strata combining the high and low extremes of both axes must contribute at least six documents in total. The pool combines these documents with filtered A-category citations, without a separate count cap for the latter.

Among candidates with valid TAC text, random selection first prioritizes negatives in the query language; any shortfall is filled from the positive document's language and then from other languages. No fixed quota is assigned to the strata. Only examples with eight valid negatives are retained.

Training also uses in-batch negatives. Let $\mathcal{I}_i$ denote documents contributed by the other training examples in the global minibatch. A similarity-based mask excludes a non-target candidate $c$ whose unscaled similarity exceeds the positive similarity by more than a margin $\delta$. Thus, with $\mathcal{M}_i=\{c\ne d_i^+:s^{(m)}_\theta(q_i,t(c))>s^{(m)}_\theta(q_i,t(d_i^+))+\delta\}$, the contrastive candidate set is $\mathcal{A}_i=(\{d_i^+\}\cup\widetilde{\mathcal{N}}_i\cup\mathcal{I}_i)\setminus\mathcal{M}_i$. The contrastive relevance objective at representation dimension $m$ is
\begin{equation}
\mathcal{L}^{X}_m=-\frac{1}{B}\sum_{i=1}^{B}\log
\frac{\exp(s^{(m)}_\theta(q_i,t(d_i^+))/\tau)}
{\sum_{c_j\in\mathcal{A}_i}\exp(s^{(m)}_\theta(q_i,t(c_j))/\tau)},
\label{eq:supervised}
\end{equation}
where $B$ is the number of queries and $t(\cdot)$ denotes the TAC representation of a candidate patent. The objective promotes the X-category target relative to the remaining comparison candidates.

Stage II directly optimizes the ranking of X-category references. Supervision is provided at the query--document level, and the model assigns a relevance score to each candidate.

\subsection{Matryoshka representation learning for dimension truncation}
\label{sec:compact}
Let $h_\theta(x)\in\mathbb{R}^{1024}$ be the encoder output for text $x$. Following Matryoshka representation learning \citep{mrl}, we define normalized prefixes
\begin{equation}
z_m(x)=\frac{h_\theta(x)_{1:m}}{\norm{h_\theta(x)_{1:m}}},
\qquad m\in\{64,1024\},
\label{eq:prefix}
\end{equation}
and use $s^{(m)}_\theta(x,y)=z_m(x)^\top z_m(y)$. The dimension-weighted supervised objective is
\begin{equation}
\mathcal{L}_{\mathrm{total}}
=\lambda\mathcal{L}^{X}_{64}+(1-\lambda)\mathcal{L}^{X}_{1024},
\qquad 0<\lambda<1.
\label{eq:mrl}
\end{equation}
Both terms update the same encoder, explicitly supervising the 64-dimensional prefix and the complete representation. The weight $\lambda$ controls their relative emphasis.

At inference, selecting and normalizing the first 64 coordinates or the complete output yields the two embedding sizes without dimension-specific retraining. Queries and candidates use the same representation dimension and normalization.

All experiments evaluate 64-dimensional representations, comparing retrieval under a common dimensionality constraint.

\section{Experiments}
\label{sec:experiments}
\subsection{Experimental setup}
The evaluation examines the retrieval effectiveness of the complete two-stage model and the contribution of weakly supervised domain adaptation relative to direct supervised fine-tuning. Aggregate results characterize performance across the collection, while analyses by language, query view, and retrieval depth examine the distribution of the observed gains.

\textbf{Evaluation collection.} Within each language and query view, all models are evaluated using the same effective test queries, candidate collection, metric definition, and aggregation procedure. The candidate collection contains approximately 300,000 patents represented by TAC (title + abstract + claim). Evaluation queries are drawn from patents filed in 2025 or later, each with at least one X-category reference recorded in an examination search report. Queries use ABS, CLAIM, DESC, or TAC. As shown in Table~\ref{tab:query-counts}, each view contains 103 effective Chinese queries and 151 effective English queries. Japanese ABS contains 125 effective queries, while Japanese CLAIM, DESC, and TAC each contain 148. The table also reports their distribution across IPC sections A--H. All scores are calculated over these effective queries using 64-dimensional embeddings. For each evaluation query, the query patent itself and all of its family members are excluded from the retrieval candidates.

\begin{table}[!htbp]
\centering
\small
\renewcommand{\arraystretch}{1.14}
\caption{Effective query counts per view, stratified by IPC section A--H. Counts on rows listing several views apply to each view separately.}
\label{tab:query-counts}
\setlength{\tabcolsep}{5pt}
\begin{tabular}{llrrrrrrrrr}
\toprule
Lang. & Query views & Total & A & B & C & D & E & F & G & H \\
\midrule
CN & All four & 103 & 9 & 11 & 13 & 11 & 13 & 18 & 12 & 16 \\
EN & All four & 151 & 20 & 22 & 17 & 16 & 21 & 19 & 16 & 20 \\
JP & ABS & 125 & 15 & 15 & 15 & 15 & 17 & 17 & 16 & 15 \\
JP & CLAIM, DESC, TAC & 148 & 18 & 18 & 21 & 17 & 19 & 20 & 17 & 18 \\
\bottomrule
\end{tabular}

\end{table}

\textbf{Training data and test exclusion.} The second-stage supervision dataset contains approximately 2.67 million examples, each consisting of a query, one positive document annotated as an X-category citation in an examination search report, and eight selected negative documents. Before both training stages, test-query patents and all of their family members are removed from the corresponding training pools.

\textbf{Baselines.} Qwen3-Embedding-0.6B is evaluated without further patent-domain adaptation, using 64-dimensional embeddings, a maximum input length of 4,096 tokens, and language-specific patent-retrieval instructions. The other baseline uses the RoBERTa architecture \citep{roberta}; it was pretrained on the full corpus of 200 million patent documents and fine-tuned for patent retrieval. Its native embedding dimensionality is 768; evaluation uses embeddings truncated to 64 dimensions. RoBERTa supports a maximum input length of 512 tokens.

\textbf{Training settings.} Training is implemented using a development version of the open-source ms-swift framework.\footnote{\url{https://github.com/modelscope/ms-swift}} Both stages optimize InfoNCE with LoRA \citep{lora} of rank 64 and $\alpha=32$. Stage I adapts all linear layers with learning rate $1\times10^{-4}$, per-device batch size 128, and gradient accumulation 1. Stage II is initialized with the parameters learned in Stage I and uses learning rate $5\times10^{-5}$, per-device batch size 16, gradient accumulation 4, in-batch negatives, and eight selected negatives per query. MRL supervises 64- and 1,024-dimensional representations with weights 0.6 and 0.4. Stage II uses contrastive temperature $\tau=0.1$ and masking margin $\delta=0.2$. Both stages use bfloat16 and a maximum sequence length of 4,096 tokens; training sequences exceeding this limit are truncated from the right. Each stage is trained for at most one epoch.

\textbf{Evaluation metric.} Hit@K measures whether at least one annotated X-category target occurs within the first K results. Hits are determined by exact document-ID matching against the annotated X-category targets, without patent-family expansion. The four Qwen-based models are evaluated at K = 1, 5, 10, 20, 50, and 100; the RoBERTa comparison is restricted to K = 100. For language $\ell$, view $v$, and IPC section $g$, let $Q_{\ell vg}$ contain the effective queries and $\mathcal{G}_{\ell v}$ the nonempty A--H sections. With $\mathcal{X}(p)$ denoting the annotated targets for source patent $p$, we compute
\begin{equation}
\begin{aligned}
h_K(q_v)&=\mathbb{I}\!\left[\mathcal{R}_K(q_v)\cap\mathcal{X}(p)\ne\varnothing\right],\\
M_{\ell v}(K)&=\frac{100}{|\mathcal{G}_{\ell v}|}
\sum_{g\in\mathcal{G}_{\ell v}}\frac{1}{|Q_{\ell vg}|}
\sum_{q_v\in Q_{\ell vg}}h_K(q_v).
\end{aligned}
\label{eq:macro-hit}
\end{equation}
At $K=100$, $M_{\ell v}$ is the IPC-section macro Hit@100 reported in Table~\ref{tab:64-detail}. The language score is $L_\ell=\frac{1}{4}\sum_{v\in\mathcal{V}}M_{\ell v}(100)$, and Overall is $\frac{1}{3}\sum_\ell L_\ell$. IPC sections receive equal weight within each language--view score, followed by equal weighting of views and languages. Scores are percentages; differences are percentage points. The same aggregation is used at all other reported cutoffs.

Equal weighting prevents more frequent IPC sections, views, or query languages from dominating the aggregate. The resulting score estimates average performance across these evaluation conditions rather than across a pooled set of queries. Disaggregated results complement this measure by characterizing variation across languages and views.

\textbf{Training comparisons and model selection.} We compare the two baselines with the weakly supervised model ($S_{\mathrm{WS}}$), direct supervised fine-tuning ($S_{\mathrm{SFT}}$), and the complete two-stage model (\model). Model selection uses a validation set of 2,000 examples partitioned from the training data pool; the model with the lowest validation loss is selected. The direct-supervision control starts from Qwen3-Embedding-0.6B and uses the same supervised data, negative construction and selection, eight negatives per query, MRL, optimization settings, input length, second-stage training budget, and validation-based selection rule as the two-stage model. This design evaluates the effect of preceding supervised retrieval training with Stage I.

\subsection{Retrieval with 64-dimensional representations}
\label{sec:64-results}
Tables~\ref{tab:64-summary} and~\ref{tab:64-detail} report the aggregate and language--view results. \model\ achieves 89.09\% overall macro Hit@100, compared with 85.89\% for direct supervised fine-tuning, an improvement of 3.19 percentage points. Scores increase in nine of the twelve language--view combinations. English and Japanese means improve by 5.34 and 4.72 points, respectively, while the Chinese mean decreases by 0.48 points.

English improves across all four views. Japanese gains are concentrated in ABS, CLAIM, and DESC, with a small decrease on TAC. Chinese results are mixed: modest gains on CLAIM and DESC are offset by decreases on ABS and TAC. Relative to RoBERTa, the two-stage model improves the Chinese and English means by 5.78 and 7.37 points, respectively. The Japanese mean remains 2.09 points lower, with higher DESC performance but lower ABS, CLAIM, and TAC scores.

The Stage I model achieves 72.06 overall, 4.22 points above Qwen3-Embedding-0.6B. Language-level changes are uneven: Chinese decreases by 2.27 points, while English and Japanese improve by 3.52 and 11.42 points, respectively. Relative to the Stage I model, the complete two-stage model improves overall Hit@100 by 17.03 points, with higher Hit@100 in all twelve language--view combinations. The complete model exceeds Qwen3-Embedding-0.6B and RoBERTa by 21.25 and 3.68 points overall, respectively.

\begin{table}[!htbp]
\centering
\small
\renewcommand{\arraystretch}{1.14}
\caption{Mean IPC-section macro Hit@100 (\%) with 64-dimensional embeddings. CN, EN, and JP average the four query views; Overall averages all twelve language--view scores. $S_{\mathrm{WS}}$ is the weakly supervised model, $S_{\mathrm{SFT}}$ is direct supervised fine-tuning, and \model\ is the complete two-stage model. Differences are computed before rounding the derived means. Bold marks the highest model score in each column.}
\label{tab:64-summary}
\setlength{\tabcolsep}{7pt}
\begin{tabular}{lrrrr}
\toprule
Model & CN & EN & JP & Overall \\
\midrule
Qwen3-Embedding-0.6B & 74.13 & 68.87 & 60.51 & 67.83 \\
RoBERTa & 85.26 & 85.68 & \textbf{85.27} & 85.40 \\
$S_{\mathrm{WS}}$ & 71.85 & 72.39 & 71.93 & 72.06 \\
$S_{\mathrm{SFT}}$ & \textbf{91.52} & 87.70 & 78.46 & 85.89 \\
\rowcolor{TableTint}
\model\ & 91.04 & \textbf{93.05} & 83.18 & \textbf{89.09} \\
\bottomrule
\end{tabular}

\end{table}

\begin{table}[!htbp]
\centering
\small
\renewcommand{\arraystretch}{1.14}
\caption{IPC-section macro Hit@100 (\%) by language and query view for the five main models. TAC denotes title + abstract + claim; Qwen3 denotes Qwen3-Embedding-0.6B. Bold marks the highest score in each row.}
\label{tab:64-detail}
\setlength{\tabcolsep}{7pt}
\begin{tabular}{llrrrrr}
\toprule
View & Lang. & Qwen3 & RoBERTa & $S_{\mathrm{WS}}$ & $S_{\mathrm{SFT}}$ & \model \\
\midrule
ABS & CN & 71.80 & 84.67 & 71.70 & \textbf{92.43} & 89.73 \\
ABS & EN & 60.97 & 84.48 & 67.72 & 83.18 & \textbf{88.16} \\
ABS & JP & 58.13 & \textbf{85.79} & 73.05 & 74.59 & 81.18 \\
\addlinespace[4pt]
CLAIM & CN & 73.88 & 89.29 & 73.88 & 91.30 & \textbf{91.38} \\
CLAIM & EN & 63.45 & 85.67 & 75.98 & 91.07 & \textbf{95.05} \\
CLAIM & JP & 59.05 & \textbf{86.37} & 69.74 & 79.55 & 84.57 \\
\addlinespace[4pt]
DESC & CN & 73.63 & 79.45 & 71.17 & 92.18 & \textbf{93.66} \\
DESC & EN & 75.27 & 80.90 & 73.53 & 88.97 & \textbf{96.96} \\
DESC & JP & 61.92 & 81.86 & 74.47 & 80.47 & \textbf{88.41} \\
\addlinespace[4pt]
TAC & CN & 77.19 & 87.63 & 70.66 & \textbf{90.15} & 89.37 \\
TAC & EN & 75.78 & 91.67 & 72.31 & 87.59 & \textbf{92.01} \\
TAC & JP & 62.92 & \textbf{87.04} & 70.45 & 79.24 & 78.55 \\
\bottomrule
\end{tabular}

\end{table}

\begin{figure}[!htbp]
\centering
\includegraphics[width=\linewidth]{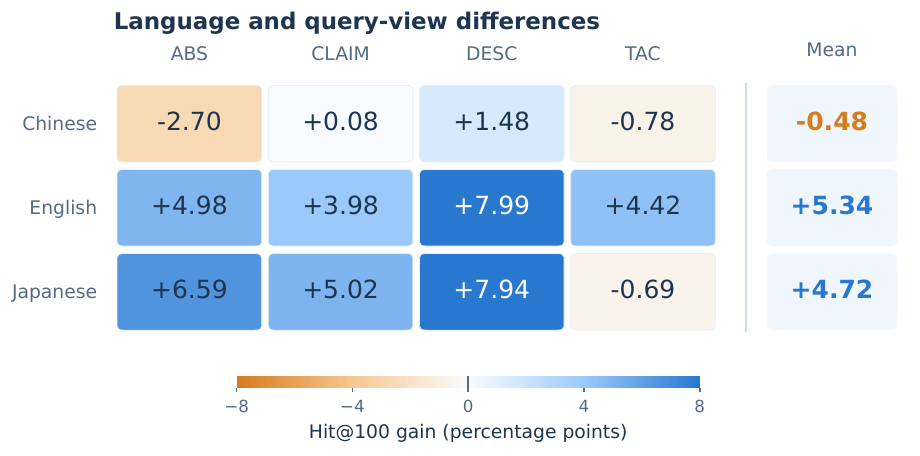}
\caption{Two-stage training gains over direct supervised fine-tuning at 64 dimensions. Each cell is the difference in IPC-section macro Hit@100 for a language--view pair; the right-hand column averages the four views within each language. All differences are percentage points, computed before rounding. The diverging color scale is symmetric about zero.}
\label{fig:results-gain}
\end{figure}

Figure~\ref{fig:results-gain} visualizes the language--view differences in Table~\ref{tab:64-detail}, together with the corresponding language averages.

\subsection{Retrieval across cutoffs}
\begin{figure}[!htbp]
\centering
\includegraphics[width=\linewidth]{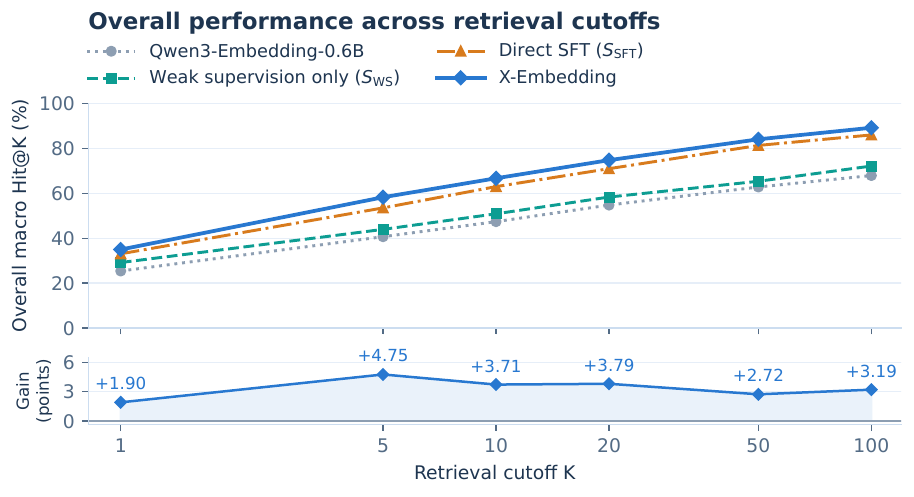}
\caption{Overall macro Hit@K at 64 dimensions (top) and two-stage gains over direct supervised fine-tuning (bottom). Scores equally weight four query views and three languages. The logarithmic horizontal axis marks all six evaluated cutoffs.}
\label{fig:results-cutoffs}
\end{figure}

Figure~\ref{fig:results-cutoffs} compares the four Qwen-based models across six retrieval cutoffs. Among the four Qwen-based models, \model\ achieves the highest overall macro Hit@K at all six cutoffs, with gains of 1.90--4.75 percentage points over direct supervised fine-tuning. Its overall macro Hit@10 and Hit@20 are 66.62\% and 74.67\%, corresponding to gains of 3.71 and 3.79 points. Stage I alone exceeds Qwen3-Embedding-0.6B at all six cutoffs but remains below both supervised models.

Table~\ref{tab:topk-control} provides the language-level comparison at K = 10 and K = 20. Two-stage training improves all three query-language groups at both cutoffs, with the largest gains for Japanese: 6.15 and 5.98 percentage points, respectively. Chinese also benefits at these cutoffs, in contrast to its slightly lower Hit@100.

\begin{table}[!htbp]
\centering
\small
\renewcommand{\arraystretch}{1.14}
\caption{Two-stage training versus direct supervised fine-tuning at Hit@10 and Hit@20 (64 dimensions, macro Hit@K in \%), using the same second-stage training configuration. Each language averages the four query views. Differences are percentage points calculated before rounding. Bold marks the higher model score at each cutoff.}
\label{tab:topk-control}
\setlength{\tabcolsep}{7pt}
\begin{tabular}{lrrrrrr}
\toprule
Language & \multicolumn{3}{c}{Hit@10} & \multicolumn{3}{c}{Hit@20} \\
\cmidrule(lr){2-4}\cmidrule(lr){5-7}
& $S_{\mathrm{SFT}}$ & \model & $\Delta$ & $S_{\mathrm{SFT}}$ & \model & $\Delta$ \\
\midrule
CN & 73.56 & \textbf{76.01} & +2.46 & 79.80 & \textbf{81.98} & +2.18 \\
EN & 64.49 & \textbf{67.02} & +2.54 & 72.18 & \textbf{75.40} & +3.22 \\
JP & 50.68 & \textbf{56.83} & +6.15 & 60.66 & \textbf{66.64} & +5.98 \\
\bottomrule
\end{tabular}

\end{table}

\section{Discussion}
\label{sec:discussion}
\subsection{The role of patent-domain adaptation}
Stage I alone changes retrieval performance unevenly across languages, yet initializing supervised fine-tuning with its parameters yields a higher aggregate score than direct supervised fine-tuning under the same second-stage conditions. The intermediate model's retrieval score therefore does not fully characterize its value as an initialization for subsequent training. The two stages target patent-content correspondence and X-category relevance, respectively; the controlled comparison supports the aggregate benefit of this training sequence on the evaluated collection.

\subsection{Language and query-view differences}
Relative to RoBERTa, the largest Hit@100 gains occur on DESC: 14.21, 16.06, and 6.55 percentage points for Chinese, English, and Japanese, respectively. The models differ in architecture, pretraining, and maximum input length: 512 tokens for RoBERTa and 4,096 for the Qwen-based models. The DESC advantage therefore cannot be attributed solely to the training strategy. The comparison with direct supervised fine-tuning, which holds the encoder architecture and second-stage training conditions fixed, more specifically evaluates the addition of patent-domain adaptation.

Language-level results describe performance on different patent query sets, so the larger English and Japanese gains do not by themselves isolate a cross-language transfer effect. Query composition also differs across Japanese views: ABS contains 125 effective queries, compared with 148 for each remaining view. The view-level scores therefore reflect both input representation and the effective query set.

Figure~\ref{fig:results-gain} and Table~\ref{tab:topk-control} show that training gains depend on query language, view, and retrieval depth. For Chinese, higher hit rates within shorter candidate lists do not coincide with an improvement in Hit@100, illustrating how different cutoffs reveal different performance changes. For Japanese, improvement over direct supervised fine-tuning coexists with a remaining gap to RoBERTa. Aggregate gains should therefore be interpreted alongside the language- and view-specific results.

\subsection{Limitations}
\label{sec:evaluation-limitations}
The evaluation is limited to one patent collection and 64-dimensional representations, with one training run per model variant. The reported differences do not establish statistical significance or variation across random seeds. The direct-supervision comparison holds the second-stage training budget fixed; adding Stage I also increases total training computation, so the results do not establish an advantage under equal total computation. Separate ablations of the two weak-supervision sources and MRL, evaluation on other collections, and retrieval at 1024 dimensions remain to be conducted. Document-level Hit@K assesses recovery of annotated references; it does not measure coverage of individual claim features, which requires claim-linked evaluation.

Evaluation is stratified by query language, without separate estimates for query--candidate language directions. X-category annotations may be incomplete, and retrieval-derived negatives without an X label can therefore include relevant documents. Test-query patents and their family members are excluded from both training stages, but their presence in the pretraining corpora of the underlying encoder and RoBERTa is unknown. These limitations constrain interpretations of cross-language transfer, pretraining overlap, and retrieval beyond the annotated reference sets.

\section{Conclusion}
\label{sec:conclusion}
We presented \model, a two-stage approach that combines weakly supervised patent-view alignment with X-category retrieval training. On the evaluated Chinese, English, and Japanese collection, the model achieves 89.09\% overall macro Hit@100 at 64 dimensions, exceeding direct supervised fine-tuning by 3.19 percentage points under the same second-stage training conditions. Overall macro Hit@10 and Hit@20 improve by 3.71 and 3.79 points, respectively, with gains in all three query-language groups at these cutoffs. At K = 100, gains are concentrated in English and Japanese, while Chinese performance decreases slightly.

The model also exceeds Qwen3-Embedding-0.6B and RoBERTa in overall macro Hit@100, although Japanese ABS, CLAIM, and TAC remain below RoBERTa. On this evaluation collection at 64 dimensions, the results show that preceding supervised retrieval training with patent-domain adaptation improves aggregate X-category retrieval, with gains varying across query languages, views, and retrieval cutoffs.

\begingroup\setlength{\baselineskip}{12pt}\setlength{\bibsep}{0pt}
\endgroup


\begin{thebibliography}{99}
\interlinepenalty=10000
\bibitem[Ayaou and Cavallucci(2025)]{ayaou2025patenteb}
Iliass Ayaou and Denis Cavallucci.
PatenTEB: A comprehensive benchmark and model family for patent text embedding.
\emph{arXiv preprint arXiv:2510.22264v1}, 2025. URL \url{https://arxiv.org/abs/2510.22264v1}.
\bibitem[Ayaou et~al.(2026)]{ayaou2025dapfam}
Iliass Ayaou, Denis Cavallucci, and Hicham Chibane.
DAPFAM: A domain-aware family-level dataset to benchmark cross domain patent retrieval.
\emph{Array}, 29:100720, 2026. doi: \href{https://doi.org/10.1016/j.array.2026.100720}{\nolinkurl{10.1016/j.array.2026.100720}}.
\bibitem[Bekamiri et~al.(2024)]{patentsberta}
Hamid Bekamiri, Daniel S. Hain, and Roman Jurowetzki.
PatentSBERTa: A deep NLP based hybrid model for patent distance and classification using augmented SBERT.
\emph{Technological Forecasting and Social Change}, 206:123536, 2024. doi: \href{https://doi.org/10.1016/j.techfore.2024.123536}{\nolinkurl{10.1016/j.techfore.2024.123536}}.
\bibitem[Daniell et~al.(2025)]{graphpatent}
Krzysztof Daniell, Igor Buzhinsky, and Sebastian Bj\"orkqvist.
Efficient patent searching using graph transformers.
In \emph{Proceedings of the 6th Workshop on Patent Text Mining and Semantic Technologies (PatentSemTech 2025)}, volume 4062 of CEUR Workshop Proceedings, pages 51--59, 2025. URL \url{https://ceur-ws.org/Vol-4062/paper5.pdf}.
\bibitem[Djemmal et~al.(2026)]{djemmal2026qaecter}
Younes Djemmal, You Zuo, Kim Gerdes, and Kirian Guiller.
Citation-driven multi-view training for patent embeddings: QaECTER and Sophia-Bench.
\emph{arXiv preprint arXiv:2604.22897v1}, 2026. URL \url{https://arxiv.org/abs/2604.22897v1}.
\bibitem[Ghosh et~al.(2024)]{ghosh2024paecter}
Mainak Ghosh, Sebastian Erhardt, Michael E. Rose, Erik Buunk, and Dietmar Harhoff.
PaECTER: Patent-level representation learning using citation-informed transformers.
\emph{arXiv preprint arXiv:2402.19411v1}, 2024. URL \url{https://arxiv.org/abs/2402.19411v1}.
\bibitem[Hu et~al.(2022)]{lora}
Edward J. Hu, Yelong Shen, Phillip Wallis, Zeyuan Allen-Zhu, Yuanzhi Li, Shean Wang, Lu Wang, and Weizhu Chen.
LoRA: Low-rank adaptation of large language models.
In \emph{International Conference on Learning Representations (ICLR 2022)}, 2022. URL \url{https://openreview.net/forum?id=nZeVKeeFYf9}.
\bibitem[Kusupati et~al.(2022)]{mrl}
Aditya Kusupati, Gantavya Bhatt, Aniket Rege, Matthew Wallingford, Aditya Sinha, Vivek Ramanujan, William Howard-Snyder, Kaifeng Chen, Sham Kakade, Prateek Jain, and Ali Farhadi.
Matryoshka representation learning.
In \emph{Advances in Neural Information Processing Systems}, volume 35, pages 30233--30249, 2022. doi: \href{https://doi.org/10.52202/068431-2192}{\nolinkurl{10.52202/068431-2192}}.
\bibitem[Li et~al.(2023)]{li2023gte}
Zehan Li, Xin Zhang, Yanzhao Zhang, Dingkun Long, Pengjun Xie, and Meishan Zhang.
Towards general text embeddings with multi-stage contrastive learning.
\emph{arXiv preprint arXiv:2308.03281}, 2023. URL \url{https://arxiv.org/abs/2308.03281}.
\bibitem[Liu et~al.(2019)]{roberta}
Yinhan Liu, Myle Ott, Naman Goyal, Jingfei Du, Mandar Joshi, Danqi Chen, Omer Levy, Mike Lewis, Luke Zettlemoyer, and Veselin Stoyanov.
RoBERTa: A robustly optimized BERT pretraining approach.
\emph{arXiv preprint arXiv:1907.11692v1}, 2019. URL \url{https://arxiv.org/abs/1907.11692v1}.
\bibitem[Risch et~al.(2021)]{patentmatch}
Julian Risch, Nicolas Alder, Christoph Hewel, and Ralf Krestel.
PatentMatch: A dataset for matching patent claims \& prior art.
In \emph{Proceedings of the 2nd Workshop on Patent Text Mining and Semantic Technologies (PatentSemTech 2021)}, volume 2909 of CEUR Workshop Proceedings, pages 40--44, 2021. URL \url{https://ceur-ws.org/Vol-2909/paper5.pdf}.
\bibitem[Robertson and Zaragoza(2009)]{bm25}
Stephen Robertson and Hugo Zaragoza.
The probabilistic relevance framework: BM25 and beyond.
\emph{Foundations and Trends in Information Retrieval}, 3(4):333--389, 2009. doi: \href{https://doi.org/10.1561/1500000019}{\nolinkurl{10.1561/1500000019}}.
\bibitem[USPTO(n.d.-a)]{uspto_anticipation}
United States Patent and Trademark Office (USPTO).
\emph{Manual of Patent Examining Procedure}, Section 2131: Anticipation---Application of 35 U.S.C. 102, n.d.-a. URL \url{https://www.uspto.gov/web/offices/pac/mpep/s2131.html}. Accessed October 19, 2026.
\bibitem[USPTO(n.d.-b)]{uspto_xcategory}
United States Patent and Trademark Office (USPTO).
\emph{Manual of Patent Examining Procedure}, Section 1844.01, VII.A.1: Particularly Relevant Documents, n.d.-b. URL \url{https://www.uspto.gov/web/offices/pac/mpep/s1844.html}. Accessed October 19, 2026.
\bibitem[van den Oord et~al.(2018)]{infonce}
Aaron van den Oord, Yazhe Li, and Oriol Vinyals.
Representation learning with contrastive predictive coding.
\emph{arXiv preprint arXiv:1807.03748v2}, 2018. URL \url{https://arxiv.org/abs/1807.03748v2}.
\bibitem[Yousefiramandi and Cooney(2026)]{yousefiramandi2026benchmarking}
Amirhossein Yousefiramandi and Ciarán Cooney.
Benchmarking patent embeddings: A multi-task evaluation of 22 models across retrieval, classification, and clustering.
\emph{arXiv preprint arXiv:2605.24297v2}, 2026. URL \url{https://arxiv.org/abs/2605.24297v2}.
\bibitem[Zhang et~al.(2025)]{qwen3embedding}
Yanzhao Zhang, Mingxin Li, Dingkun Long, Xin Zhang, Huan Lin, Baosong Yang, Pengjun Xie, An Yang, Dayiheng Liu, Junyang Lin, Fei Huang, and Jingren Zhou.
Qwen3 Embedding: Advancing text embedding and reranking through foundation models.
\emph{arXiv preprint arXiv:2506.05176}, 2025. URL \url{https://arxiv.org/abs/2506.05176}.
\bibitem[Zuo et~al.(2026)]{zuo2025selfsupervision}
You Zuo, Kim Gerdes, Éric de la Clergerie, and Benoît Sagot.
Patent representation learning via self-supervision.
In \emph{Proceedings of the 2026 International ACM SIGIR Conference on Innovative Concepts and Theories in Information Retrieval (ICTIR 2026)}, pages 436--445, 2026. doi: \href{https://doi.org/10.1145/3805713.3820439}{\nolinkurl{10.1145/3805713.3820439}}. Corrected version: URL \url{https://arxiv.org/abs/2511.10657v3}.
\end{thebibliography}
\end{document}